\documentclass[sigconf]{acmart}  

\usepackage{color}
\usepackage{balance}
\usepackage{algorithm} 
\usepackage{algpseudocode}

\def\BibTeX{{\rm B\kern-.05em{\sc i\kern-.025em b}\kern-.08emT\kern-.1667em\lower.7ex\hbox{E}\kern-.125emX}}

\renewcommand\footnotetextcopyrightpermission[1]{}
\copyrightyear{2023}
\acmYear{2023}
\setcopyright{acmlicensed}
\acmConference[xxxx 2023]{xxxx}{xxxx.xxxx, 2023}{xxxx, xxxx}
\acmBooktitle{xxxx, xxxx. xxxx, 2023,xxxx, xxxx}

\begin{document}

%
\title{FedCIP: Federated Client Intellectual Property Protection with Traitor Tracking
}

%
\author{Junchuan Liang}
\email{liangjunchuan@stu.sicnu.edu.cn}
\author{Rong Wang}
\authornotemark[1]
\email{rwang@sicnu.edu.cn}
\affiliation{%
  \institution{School of Computer Science, Sichuan Normal University}
  \streetaddress{Chenlong Street 1819}
  \city{Chengdu}
  \state{China}
  \postcode{610101}
}







%

%
\begin{abstract}
Federated learning is an emerging privacy-preserving distributed machine learning that enables multiple parties to collaboratively learn a shared model while keeping each party's data private. However, federated learning faces two main problems: semi-honest server privacy inference attacks and malicious client-side model theft. To address privacy inference attacks, parameter-based encrypted federated learning secure aggregation can be used. To address model theft, a watermark-based intellectual property protection scheme can verify model ownership. Although watermark-based intellectual property protection schemes can help verify model ownership, they are not sufficient to address the issue of continuous model theft by uncaught malicious clients in federated learning. Existing IP protection schemes that have the ability to track traitors are also not compatible with federated learning security aggregation. Thus, in this paper, we propose Federated Client-side Intellectual Property Protection (FedCIP), which is compatible with federated learning security aggregation and has the ability to track traitors. To the best of our knowledge, this is the first IP protection scheme in federated learning that is compatible with secure aggregation and has tracking capabilities.
\end{abstract}

%


%

%

%
\maketitle

%
\section{Introduction}
Federated learning (FL) \cite{34} has emerged as a popular framework for training deep learning models without sharing private data, with applications in medical image analysis \cite{35,36}, recommendation systems \cite{37}, and object detection \cite{38}. Federated learning provides the ability to protect data privacy while still utilizing large amounts of private data to generate accurate and reliable models. However, federated learning also poses risks such as privacy inference attacks \cite{21,22,23,24} and model theft \cite{30}. In privacy inference attacks, attackers are often honest but curious servers or malicious attackers with direct access to client update parameters. In model theft, attackers are often malicious clients who compromise the security of the federated learning system, posing a significant threat to the privacy-preserving features and outcomes of federated learning.

To mitigate the risk of privacy inference attacks, various secure aggregation methods for federated learning have been proposed, including parameter encryption techniques \cite{26} and differential privacy \cite{27,28,29}. Parameter encryption allows the server to perform aggregation computation while keeping the real parameters inaccessible, effectively protecting model parameters from disclosure. Differential privacy involves adding a certain amount of noise to the model parameters according to the differential privacy budget, making the original data more difficult to infer. The core of all these approaches is to prevent the server from accessing the complete real model parameter information, ensuring that the server cannot perform any operations (such as embedding backdoors to the global model) on the model parameters other than necessary computation.

To combat model stealing attacks in federated learning, watermarking \cite{1,2,3,4,5,6} techniques have been proposed as an effective solution. These techniques can be divided into two categories: client-side watermarking \cite{15,16} and server-side watermarking \cite{17,18,19}. Client-side watermarking involves injecting a watermark into the model during the local training process of the client, while server-side watermarking involves embedding a slightly different watermark into each model distributed to the client at each model aggregation. Both approaches effectively verify the ownership of suspicious models. But in federated learning, it is not enough to just verify ownership. This is because traitors are able to consistently steal models if they are not caught. However, existing methods that can track traitors are not compatible with federated learning security aggregation. Methods that are compatible with federated learning security aggregation again do not have tracking capabilities. A solution with both capabilities means that the server distributes the same model to each client and then finds who stealing the model. Even previous research has suggested that if the server distributes the same model to each participant, this can lead to difficulties in tracking traitors \cite{18}.

We propose a novel approach to solve this problem which is called FedCIP (Federated Client Intellectual Property Protection). FedCIP has the ability to track traitors while being compatible with federated learning security aggregation. In our attack scenario, there are two main adversaries: the honest but curious server, and the persistent model stealing client. To be compatible with the Federated Learning Security Aggregation approach, FedCIP servers cannot do anything to the model beyond the necessary calculations. Then, in order to track traitors in the client, FedCIP employs a fast and replaceable client-side watermarking technique along with client selection to achieve this. 

FedCIP first utilizes replaceable client-side watermarking techniques to update federated learning models at each round or several rounds, allowing for unique watermarks in each model. By comparing these watermarks, we can identify the rounds in which the model was stolen. The server selection aggregation of federated learning is then utilized to select only certain clients who participated in these rounds, which results in a set of suspicious traitors. Finally, multiple sets of potential traitors obtained from suspicious models are examined to identify the same client among them, who is the traitor. FedCIP only requires the client to embed the watermark, and the server does not make any additional modifications to the model, making it compatible with federated learning security aggregation methods such as homomorphic encryption. Furthermore, it has the ability to track traitors who continuously steal the model on the client side.

Moreover, FedCIP also has the fundamental capabilities of an IP protection scheme. These capabilities include ownership verification, resistance to pruning attacks, resistance to fine-tuning attacks, etc. We compare our program (FedCIP) with the existing Federated Learning IP protection program in Table \ref{tab:1}.

\begin{table*}[htbp]
  \centering
  \caption{Comparison of the federated learning IP protection program.}
  \label{tab:1}
  \begin{tabular}{lcccc}
    \toprule
    Method&Pruning resistant&Fine-tuning resistant&Secure FL&Traitor tracking\\
    \midrule
    Merkle-Sign \cite{19} &Yes&Yes&No&Yes\\
    WAFFLE \cite{17} &Yes&Yes&No&No\\
    FedIPR \cite{15} &Yes&Yes&Yes&No\\
    FedTracker \cite{18} &Yes&Yes&No&Yes\\
    Watermarking in Secure FL \cite{16} &Yes&Yes&Yes&No\\
    FedTracker \cite{18} &Yes&Yes&No&Yes\\
    FedCIP &Yes&Yes&Yes&Yes\\
  \bottomrule
\end{tabular}
\end{table*}

The main contributions of this article are the following three points.

\begin{itemize}
\item We have introduced the FedCIP framework, which is a novel approach for protecting the intellectual property of federated learning clients, capable of tracking traitors while compatible with federated learning security aggregation.

\item To facilitate the implementation of FedCIP, we have proposed a Fast Federated Watermarking algorithm ($FFWM1$, $FFWM2$), which allows federated learning clients to replace watermarks.

\item Experiments demonstrated the effectiveness and robustness of our approach through extensive experiments. Additionally, we have conducted a thorough theoretical analysis of the number of traitors leaking the model using our method and the probability of them being caught.
\end{itemize}

\section{Backgrounds}

\subsection{Privacy inference attack}
A privacy inference attack \cite{25} refers to an attack in which an adversary tries to infer the original data by analyzing multiple rounds of gradient values during the training process in FL. The attacker in a privacy inference attack is usually an honest but curious server or another malicious individual who has access to model updates. Such attacks pose a significant threat to the very essence of federated learning, which is to protect the privacy of participants' data.

Zhu et al. \cite{20} proposed  Deep Leakage from Gradients (DLG) algorithm which utilizes leaked gradient information to make inferences about the raw data. Building on this work, Zhao et al. \cite{21} present the iDLG algorithm, which offers improved inference under the same conditions. Additionally, Yin et al. \cite{22} demonstrate high-volume privacy data recovery, while Ren et al. \cite{23} and Wang et al. \cite{24} propose privacy inference using Generative Adversarial Network (GAN) based on client-side gradient information.

\subsection{Defenses against privacy inference attack}

To defend against privacy inference attacks in federated learning, the main strategies are based on cryptographic principles such as homomorphic encryption \cite{26} and differential privacy \cite{27,28,29}.

\textbf{Homomorphic encryption.} Homomorphic encryption \cite{26} is a powerful cryptographic technique that allows computations to be performed on encrypted data. This means that the server can perform computations on models while protecting its confidentiality. In federated learning, homomorphic encryption is used to further enhance the privacy protection of participants' model. This approach enables the server to compute while protecting the real model parameters. This greatly reduces the risk of data leakage during the federated learning process. 

\textbf{Differential privacy.} Differential privacy \cite{27,28,29} is a technique that involves adding noise to data in order to protect the privacy of the data. Unlike indiscriminately adding noise, differential privacy involves adding noise with specific size and distribution requirements. The key characteristic of differential privacy is that it enables the protection of the privacy of the data while maintaining the usability of the data.

\subsection{Model theft}

Federated learning is vulnerable to free-rider attackers and illegal copy \cite{30}. Due to the need to protect the privacy of client data and processes are not visible to others in the federated learning process. This provides an opportunity for attackers to masquerade as benign participants in federated learning and steal models, and continuously leak them. These attackers also try to disable any potential watermarking in the model through watermarking removal methods such as model \textbf{fine-tuning} \cite{32} and \textbf{pruning} \cite{12,31}. Model fine-tuning involves the attacker training the model for a certain number of rounds using local data instead of directly leaking the stolen model. By changing the model weight parameters through this fine-tuning method, the potential watermarking can be destroyed while ensuring the model's availability. Model pruning, on the other hand, is a technique used by attackers to prune the unimportant weight nodes in the stolen model to remove the potential watermarking.

\subsection{Defenses against model theft}

\subsubsection{DNN watermarking}

Watermarking \cite{1,2,3,4} is a technique in Deep Neural Networks (DNN) to safeguard intellectual property, and can be classified into backdoor-based \cite{5,6,7,8,9} and parameter-based \cite{10,11,12,13,14} methods.

\textbf{Backdoor-based watermarks.} The backdoor-based \cite{5,6,7,8,9} approach was first introduced as an attack method \cite{1,2,3}, but it is also being used for intellectual property protection \cite{4,5,6} due to its ability to directly control the output of the model using specific triggers without modifying the model parameters. This method has a minimal impact on the model's primary task and is thus considered a popular black-box approach \cite{7,8,9}.

\textbf{Parameter-based watermarks.} One of the approaches to embedding watermarks in deep neural networks is the parameter-based method \cite{10,11,12,13,14}, which is a white-box technique that involves modifying the model parameters to embed the watermarking. This method allows for quick watermarking injection but requires direct access to the model parameters. In the parameter-based watermarking approach, the watermarking is represented as an N-bit string and is directly embedded into the model parameters $W$.

\subsubsection{FL watermarking}

Relying on the (DNN) intellectual property protection approach is inadequate for the safeguarding of intellectual property in FL models. Firstly, this is because multiple clients participate in FL and pose potential threats to the model's security. Consequently, intellectual property protection in FL must continue throughout the entire training process. Secondly, the model aggregation process in FL may weaken watermarks that lack robustness. Lastly, tracking traitors among participating clients in FL and excluding them is crucial to better safeguard the intellectual property rights of other participants. There are two main areas of research on intellectual property protection in the field of FL.

\textbf{Client-based method.} One of the most widely used methods for protecting intellectual property in federated learning is client-side watermarking \cite{15,16}. Among these methods, a client-side watermarking scheme that can satisfy federated learning security aggregation was proposed in \cite{15}. This approach allows for ownership verification of leaked models and ensures the robustness of the watermarking. Similarly, Yang et al. \cite{16} proposed a new scheme for protecting intellectual property in federated learning scenarios, assuming the existence of special clients and their watermarking of models, while also allowing for verification of the ownership of leaked models. However, neither of these methods has the capability to track traitors in federated learning. This leads to the issue that the leaker can leak with impunity because they will not be caught.

\textbf{Server-based method.} This approach begins with the server and uses watermarking injection on the model through the server. Tekgul et al. \cite{17} proposed a server backdoor approach, which can generate backdoor triggers without training data and can be effectively triggered by the model to achieve intellectual property protection. Shao et al. \cite{18} proposed a server watermarking method, which can prove the ownership of the model while tracking traitor among the clients through a unique watermarking for each client. A similar approach for model ownership verification and traitor tracking through the client is presented in \cite{19}. In addition, Li et al. \cite{19} proposed a design scheme for recovering the identity of watermarked lost participants. However, these server-side watermarking methods are at risk of privacy inference attacks as they fail to meet the requirements of federated learning security aggregation schemes.

\section{Threat model and security requirements}

\subsection{Threat model}

There are two main threat models that we face. The first one is the honest but curious server. The second one is model theft in the client.

The honest but curious server is a server that performs honest computation as requested but is curious about the client's private data. This curiosity may come from the server itself or from a potential attacker hidden in the server. Curious servers can pose a threat to the privacy and security of federated learning.

Model theft masquerades as a benign client to join FL. The model thief behaves like a benign client except for leaking models. And since federated learning is an ongoing, long-term process, multiple global models are generated during the process. The model thief can continuously steal and leaks the global model of federated learning.

\subsection{Security requirements}

To address the threats mentioned above, we suggest that an intellectual property protection solution in a federated learning setting should satisfy the following two aspects. On the one hand, this solution is able to resist honest and curious servers. On the other hand, it can resist consistent model theft. For the latter, in addition to verifying the ownership of suspicious models, it should be able to track traitors in the client and address model theft at the root. Therefore, we summarize the security requirements of the federated learning IP protection scheme in the following points.

\textbf{Secure aggregation.} The approach should satisfy the Federated Learning Security Agency's requirements. It must ensure that the server cannot directly modify the parameters inside the model.

\textbf{Traitor tracking.} In federated learning, verification of suspicious model ownership alone does not prevent model theft. Therefore, it is necessary to track down traitors who persistently steal models.

\textbf{Fidelity.} The approach should ensure that the model performance does not impact significantly after watermarking.

\textbf{Robustness.} The approach should be resistant to watermarking removal techniques such as model fine-tuning and pruning to maintain ownership of the verification capability.

\textbf{Reliability.} The added watermarking should have a high recognition rate under the verification mechanism and a low recognition rate for non-watermarked models.

\textbf{Capacity.} A reliable watermarking mechanism should have a flexible capacity to bury a varying number of watermarking bits based on the model's size, ensuring a good recognition rate.

\textbf{Independence.} The watermarking generation should not rely on or modify the original training data to ensure the integrity of the original data.

\section{FedCIP framework}
\subsection{Overall design}



\begin{figure*}[htbp]
  \centering
  \includegraphics[width=\linewidth]{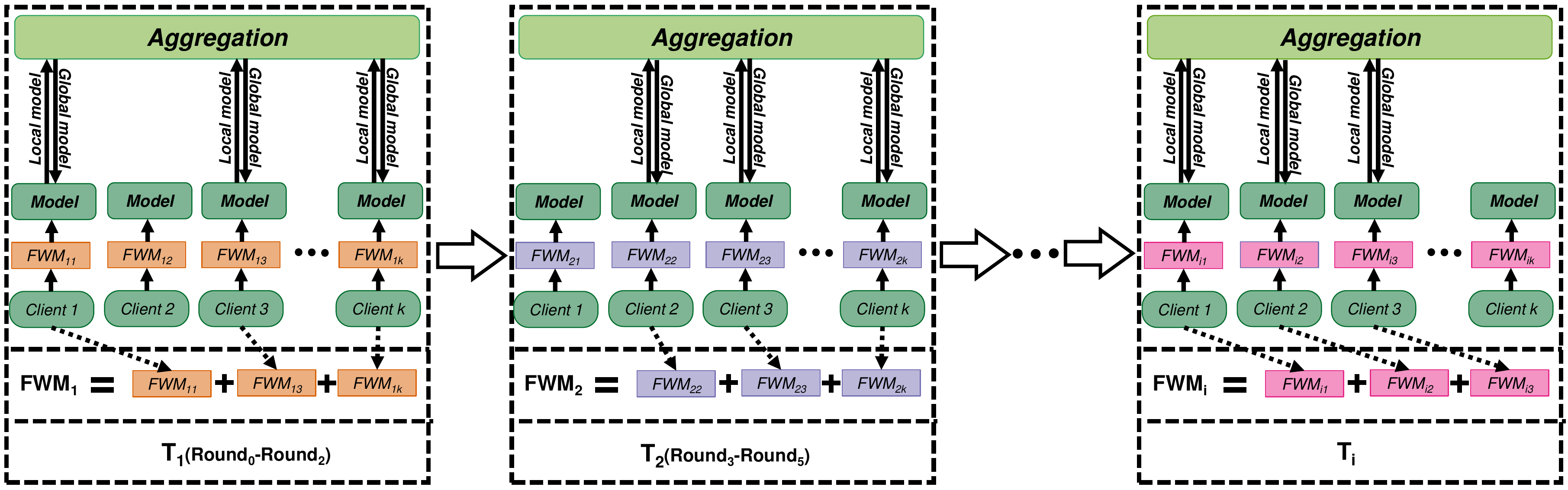}
  \caption{FedCIP watermark embedding training process.}
  \Description{FedCIP watermark embedding training process.}
  \label{fig1}
\end{figure*}

\begin{table}[htbp]
  \caption{provides the notation used in this section.}
  \label{tab:2}
  \begin{tabular}{cl}
    \toprule
    Parametric&Description\\
    \midrule
    $\alpha$	&The watermark embedding power parameter \\
            &when the round is not the first in a cycle\\
    $\beta$	&The watermark embedding power parameter\\
            &when the round is the first in a cycle\\
    $K$	&Total number of clients participating in FL\\
    $C$	&Coefficient of the number of client aggregations\\
        &selected by the server per round\\
    $T_i$	&Watermark update cycle; $T_i$=1, 2, 3...round\\
    $\eta$	&Purning rate\\
    $FWM_i$	&Watermark embedded by all client in $T_i$\\
    $FWM_w$	&Watermark extracted from the model\\
    $FWM_{ik}$	&Watermark embedded by the $kth$ client in $T_i$\\
    $FWM_{wk}$	&Watermark extracted from the $kth$ client\\
    $M$	&The model before the watermark was embedded\\
    $\tilde{M}$	&The model after being embedded with watermark\\
    $W$	&Weights before being selected by the client to \\
            &embed the watermark\\
    $\tilde{W}$	&The weight after being selected by the client to \\
            &embed the watermark\\
    $WMdacc$	&Accuracy of watermark detection\\
    $N$	&Watermark length\\
    $N_t$	&Number of clients that may be traitors\\
  \bottomrule
\end{tabular}
\end{table}

FedCIP is designed to ensure resistance against honest and curious servers while also having the ability to track traitors in the client. To achieve this, FedCIP introduces the concept of a training cycle $T_i$ in the federated learning training process shown in Figure \ref{fig1}. The training cycle $T_i$ is defined as a $T_i$ containing multiple rounds. For example, $T_i$=3, which means we consider 3 rounds as one cycle. Different cycles embed different watermarks, and the same watermark is used in the same cycle. In the usual federated learning, the server selects different clients in each round to participate in aggregation. However, in FedCIP, the server selects clients in $T_i$. That is, within a $T_i$, the clients participating in the federated aggregation are all the same. But different $T_i$'s participating are different. In particular, when the $T_i$=1 round, watermark replacement is performed for each round. Then, at each $T_i$, the model is embedded with the corresponding unique Federated Watermark ($FWM_i$). The $FWM_i$ is embedded by clients selected by the server to participate in federated aggregation.

In addition, the server will do the work of the region division to prevent client watermark conflicts and logging. In the first model distribution phase, the server will divide the model into K regions to be watermarked according to the number of clients K. For example, the model parameters have a total of 1,000,000 weight parameters, and the number of clients K=10. Then the server will divide all parameters into 10 parameter regions and randomly assign them to K clients. Each client will be given a region with 100,000 parameters to be watermarked and used to embed the watermark. The clients will embed the watermark in their respective regions. This area is mainly divided to prevent watermark conflicts. Secondly, the server will record the number of clients participating in federated learning in each cycle, which is used to verify the model watermark and find the traitor. Moreover, FedCIP utilizes parameter-based watermarking techniques on the client side as the foundation, incorporating server selection and Fast Federated Watermarking.

\subsection{Client-side federated watermark embedding}
\subsubsection{Cycle division}

The watermark embedding process of the FedCIP algorithm is divided into two main parts. The first part involves partitioning the training rounds by $T_i$ and then performing the watermark embedding algorithm within $T_i$. There can be multiple rounds inside each $T_i$, where the first round use Fast Federated Watermark 1 ($FFWM1$) and the rest use Fast Federated Watermark 2 ($FFWM2$).

\renewcommand{\algorithmicrequire}{\textbf{Input:}}
\renewcommand{\algorithmicensure}{\textbf{Output:}}

\renewcommand{\thealgorithm}{1} 
    \begin{algorithm}
    \label{al1}
        \caption{FedCIP  watermark embedding algorithm} 
        \begin{algorithmic}[1] 
            \Require Total number of training rounds $rounds$, watermark cycle $T_i$, watermark $FWM_i$, clients participating in federated learning $CK$, global model $M$
            \Ensure Watermarked model $\tilde{M}$.

            \For{$round$ in range($rounds$)}
                \If{$round\ mod\ T_i == 0$}
                    \For{$k$ in $CK$}
                        \State $local_{model}\leftarrow$ deepcopy($M$);
                        \State Train($local_{model}$);
                         \State $FFWM1$;
                    \EndFor
                \Else   
                    \For{$k$ in $CK$}
                        \State $local_{mode}\leftarrow$ deepcopy($M$);
                        \State Train($local_{model}$);
                        \State $FFWM2$;
                    \EndFor
                \EndIf 
            \EndFor
            \State Execute Federated Aggregation Algorithm FedAvg;
            \State $\tilde{M} \leftarrow$ Watermark($M, FWM_i$);
            \State \Return{$\tilde{M}$}
        \label{al1}    
        \end{algorithmic}
    \label{al1}    
    \end{algorithm}

Algorithm \ref{al1} is used for cycle division in FedCIP. The algorithm randomly selects a certain percentage of clients to participate in the current $T_i$ for watermark embedding. The watermark is then embedded in the selected clients' models using the Fast Federated Watermark algorithm. After $T_i$ rounds of federated learning, the watermark is replaced with a new one. The new watermark replaces the previous one completely and is embedded in the models of the selected clients for the next $T_{i+1}$ cycle. This process continues until the end of the training process. It is important to note that in each $T_i$ cycle, the same watermark is embedded in the models of all the selected clients. However, the watermark is different in each $T_i$ cycle.

The algorithm operates in multiple training rounds, with the watermark applied at specific intervals defined by the watermark cycle $T_i$. The input parameters for the algorithm include the total number of training rounds, watermark cycle $T_i$, watermark $FWM_i$, participating clients $CK$, and the initial global model $M$.

After the above process, in each cycle, $T_i$, the global model of federated learning is embedded in the corresponding watermark $FWM_i$

\subsubsection{FFWM1}

The $FFWM1$ algorithm is executed at only the first round of each $T_i$. The purpose of $FFWM1$ is to quickly replace the original watermark with a new one. the $FFWM1$ algorithm is as follows.


\begin{equation}
\tilde{W}=	W+\beta \times(FWM_{ik}-FWM_{Wk})
\end{equation}

\begin{equation}
FWM_{Wk}=
\begin{cases}
1& (if\quad W>0)\\
-1& (if\quad W<0)
\end{cases}
\label{eq2}
\end{equation}

$\tilde{W}$ is the weight after the client embeds the watermark; $W$ is the weight before the watermark is embedded; $\beta$ is the watermark embedding power; $FWM_{ik}$ is the watermark that will be embedded in that cycle by $kth$ client. $FWM_{Wk}$ is the watermark extracted from the weight before the watermark is embedded in the $kth$ client model. $FWM_{Wk}$ is calculated as shown in Equation \ref{eq2}. The algorithm takes as input the weight vector $W$ used by the client to embed the watermark, which can be considered as a vector, and the watermark $FWM_{Wk}$ used in the current cycle, which is a binary array of equal length to $W$ containing only 1 and -1. The output is the weight vector $\tilde{W}$ after embedding the watermark.

The algorithm first initializes the watermark $FWM_{Wk}$ for the initial weight vector $W$. If the weight vector $W$ is greater than 0, the corresponding element of $FWM_{Wk}$ is set to 1, otherwise it is set to -1. Then, the algorithm calculates the weight vector $\tilde{W}$ using the formula $\tilde{W}  =	W+\beta \times(FWM_{ik}-FWM_{Wk})$, where $\beta$ is the power factor of the embedded watermark.

Note that the position of $W$ is selected by the client during the first training, and the position remains unchanged thereafter. Furthermore, the binary array $FWM_{W}$ is used to determine which weights need to be replaced with the new watermark. Only those weights that are different from the previous watermark need to be replaced, as this can reduce the changes to the model.

\subsubsection{FFWM2}

Algorithm $FFWM2$ is used for embedding watermarks in non-first rounds when $T_i$ is greater than 1. The $FFWM2$ algorithm is as follows.

\begin{equation}
    \tilde{W}=\alpha \times FWM_{ik}
\end{equation}

Where $\tilde{W}$ is the weight after embedding the watermark. $\alpha$ is the watermark strength factor, and $FWM_{ik}$ is the current cycle watermark.

Two reasons exist for using a different watermarking method than the first round. First, the first round requires rapid experimental watermark replacement, so the change in parameters is relatively large. However, if the case where $T_i$ is greater than 1, the subsequent rounds do not need a large number of parameter changes. Second, for the first round, excessive or small parameter changes can be constrained to the $FWM_{ik}$ neighborhood, making the watermark more robust.

\begin{figure*}[htbp]
  \centering
  \includegraphics[width=0.8\linewidth]{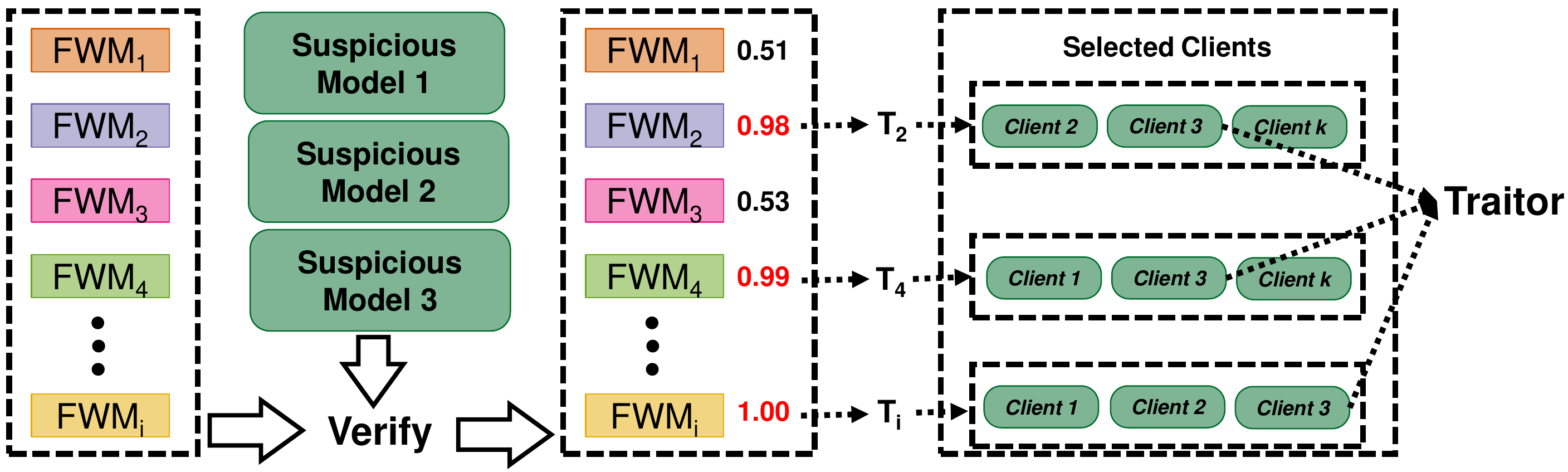}
  \caption{Watermark verification and traitor tracking.}
  \Description{Watermark verification and traitor tracking.}
  \label{fig12}
\end{figure*}

\subsection{Ownership verification}

\subsubsection{Federated watermark}

The ownership of the model is verified using the federated watermark. The federated watermark is the sum of all the watermarks of the participating aggregation clients in the $T_i$ cycle. It is defined as follows.

\begin{equation}
    FWM_i= Concat [FWM_{ik} (k=1,2,3,...,k \in CK )]
\end{equation}

$FWM_i$ is the federated watermark of $T_i$ cycle and $FWM_{ik}$ is the $kth$ client's watermark. For example, in the $T_i$ cycle, the server selects a total of 8 clients to participate in the federated learning aggregation, and the length of each client watermark is 10 bits, so the federated watermark is 80 bits.

\subsubsection{Suspicious model watermark extraction}

Next, we will extract the watermark from the suspicious model by determining the corresponding weight values based on the 
$FWM_w$ calculated in Equation \ref{eq5}. The segmentation threshold value of 0 is utilized during the extraction process. The extraction method can be described as follows.

\begin{equation}
    FWM_{w}=
\begin{cases}
1& (if\quad W>0)\\
-1& (if\quad W<0)
\end{cases}
\label{eq5}
\end{equation}

$FWM_w$ is the watermark extracted from the suspicious model. 

\subsubsection{Watermark detection}

We use the Hamming distance as a measure for watermark comparison. The Hamming distance can be used to measure the difference between two binary arrays. We define the watermark detection accuracy as $WMdacc$ and the expression is as follows.

\begin{equation}
    WMdacc = 1-1/N\times \mathscr{H}(FWM_i,FWM_{w}) \quad (N\ge 1)
\end{equation}

$\mathscr{H}$ is the Hamming distance, $FWM_i$ is the federated watermark, and $FWM_{w}$ is the suspicious model extracted watermark, $N$ is the length of the federated watermark.

\subsection{Traitor tracking}

The core intuition of traitor tracking is that if a client skips training in a cycle, their local model will lack the watermark generated by the participating clients in that cycle's aggregation. Each federated watermark is unique to the cycle, allowing us to identify which cycle a leaked model originated from by analyzing its watermarks. From there, we can obtain the set of potential traitors for that specific cycle. Finally, multiple sets of potential traitors were obtained from
suspicious models and the same client in all sets is the traitor as Figure \ref{fig12}.

We define the server's selection rate of clients per cycle as $C$, which means that $CK$ clients are selected per cycle (where $K$ is the total number of participating clients in federated learning). Assuming that the traitor leaks the model $n$ times, we can obtain n set $C_p$ of potential traitors, each of which contains the traitor. The problem we need to solve is determining the minimum number of leaks required before we can identify the traitor in $C_p$. We assume that the expected number of potential traitors in the set $C_{p1}$ obtained from each leak is $N_t$. If there is only one leak, the expected number of potential traitors is simply the number of clients in $C_{p1}$. However, if there is a second leak and we obtain a second set of potential traitors $C_{p2}$, the expected number of potential traitors becomes the number of $(C_{p1} \cap C_{p2})$. We can uniquely identify the traitor when this expected number $N_t$ is less than one. Therefore, the required number of leaks is

\begin{equation}
    N_t =K \times C^{n} \le 1
\end{equation}

In this scenario, $K$ represents the total number of clients participating in federated learning, while $C$ denotes the server's selection rate. The traitor leaks the model $n$ times. For example, if there are 10 clients participating in federated learning with a server selection rate of 0.5 (meaning models from 5 clients are aggregated each time), then the traitor will be caught on average after stealing  the model 4 times. From the traitor's perspective, the probability of being caught when leaking the model for the first time is 0.5, increasing to 0.75 for the second leak and 0.875 for the third leak. By using replaceable watermarking (which does not require additional storage) and logging the clients involved in each cycle, our approach can identify traitors with a higher risk of being caught when stealing the model.

\section{Experiments}

\subsection{Evaluation metrics}

We use the following two evaluation metrics to assess our approach.

\textbf{Model accuracy.} The accuracy of the network model on the test set during training. It reflects the basic performance of the model and will be used as a measure of model performance.

\textbf{Watermark detection accuracy.} Watermark detection accuracy ($WMdacc$) is defined in detail in section 4.3.3 of the paper. The $WMdacc$ measures the difference between the watermark before the client embedding and the watermark extracted from the suspicious model. A value of 1 indicates that the two are identical, and this metric will be used to evaluate the effectiveness of watermark detection.

\subsection{Experiment setting}

\subsubsection{Dataset}

We use the following three data sets for our experiments.

\textbf{MNIST.} The MNIST \cite{40} consists of 70,000 grayscale images of handwritten digits, from 0 to 9. Each image is 28x28 pixels in size, and the intensity of each pixel ranges from 0 (white) to 255 (black). The dataset is split into two parts: 60,000 training images and 10,000 test images. This division allows researchers and practitioners to train their algorithms on the training set and evaluate their performance on the test set.

\textbf{FashionMNIST.} FashionMNIST \cite{41} is a dataset of Zalando's article images. The dataset consists of 70,000 grayscale images, each with a resolution of 28x28 pixels. There are 10 different classes of fashion items, with each class having 7,000 images. The 10 classes are T-shirts/tops, Trousers, and Pullovers... FashionMNIST is divided into two parts: a training set of 60,000 images, used for training machine learning models, and a test set of 10,000 images, used for evaluating the performance of trained models.

\textbf{CIFAR-10.} The CIFAR-10 \cite{42} dataset (Canadian Institute For Advanced Research) is a widely used dataset for image classification in the field of machine learning and computer vision. The dataset consists of 60,000 32x32 color images, divided into 10 classes such as Airplane, Automobile, Bird, and so on. Each class contains 6,000 images, and the dataset has a total of 50,000 training images and 10,000 test images.

The above three datasets are used in experiments using Independent Identical Distribution (IID) and Non-Independent Identical Distribution (non-IID). The non-IID setting uses label-based non-IID. i.e., the training data is labeled incompletely for each client. In our non-IID setting, each client has only random data with 2-5 labels.

\subsubsection{Network model}

We used the following three network models for our experiments.

\textbf{MLP.} MLP stands for multilayer perceptron, which is a neural network model used in supervised learning tasks. MLP is a type of feedforward neural network, which means that information flows through the network in one direction, from the input layer to the output layer, without any feedback loops. Our model has 2 hidden layers, each with 200 units and ReLU activation functions. The input layer has 784 nodes (28x28 pixels) corresponding to the size of the images in the dataset, and the output layer has 10 nodes corresponding to the 10 possible digits (0-9) that can be present in the images. This model has a total of 199,210 trainable parameters, which includes the weights and biases of the 3 fully connected layers.
    
\textbf{CNN.} The model consists of two convolutional layers (conv1 and conv2) with kernel sizes of 5x5, followed by max-pooling layers (pool) with a kernel size of 2x2. The convolutional layers extract features from the input images, and the max pooling layers reduce the size of the feature maps to increase computational efficiency. The ReLU activation function is applied to the output of each convolutional layer. This model has a total of 1,668,426 trainable parameters, which include the weights and biases of the convolutional layers, the weights and biases of the fully connected layer, and the weights and biases of the output layer.
    
\textbf{ResNet18.} The ResNet18 \cite{43} architecture consists of 18 layers and uses residual connections to allow for deeper networks to be trained without encountering the vanishing gradient problem. The model is divided into four groups of layers, each consisting of multiple BasicBlock layers, which are defined in the code above. Each BasicBlock consists of two convolutional layers followed by batch normalization and a ReLU activation function and a shortcut connection that allows the input to bypass the convolutional layers. The ResNet18 architecture has a total of 11,173,962 trainable parameters, which include the weights and biases of the convolutional layers and fully connected layers.

\subsubsection{Federated learning settings}

The federated learning aggregation method in this paper uses federated average (FedAvg) \cite{34}. The default total number of clients participating in federated learning is 10. The server selection rate in each round is 0.5. i.e., the server selects 5 clients for model aggregation and watermark embedding in each round. The learning rate is 0.01. The number of local training rounds per client is 2.

\subsection{Experimental results}

\subsubsection{Fidelity}

\begin{figure}[h]
  \centering
  \includegraphics[width=\linewidth]{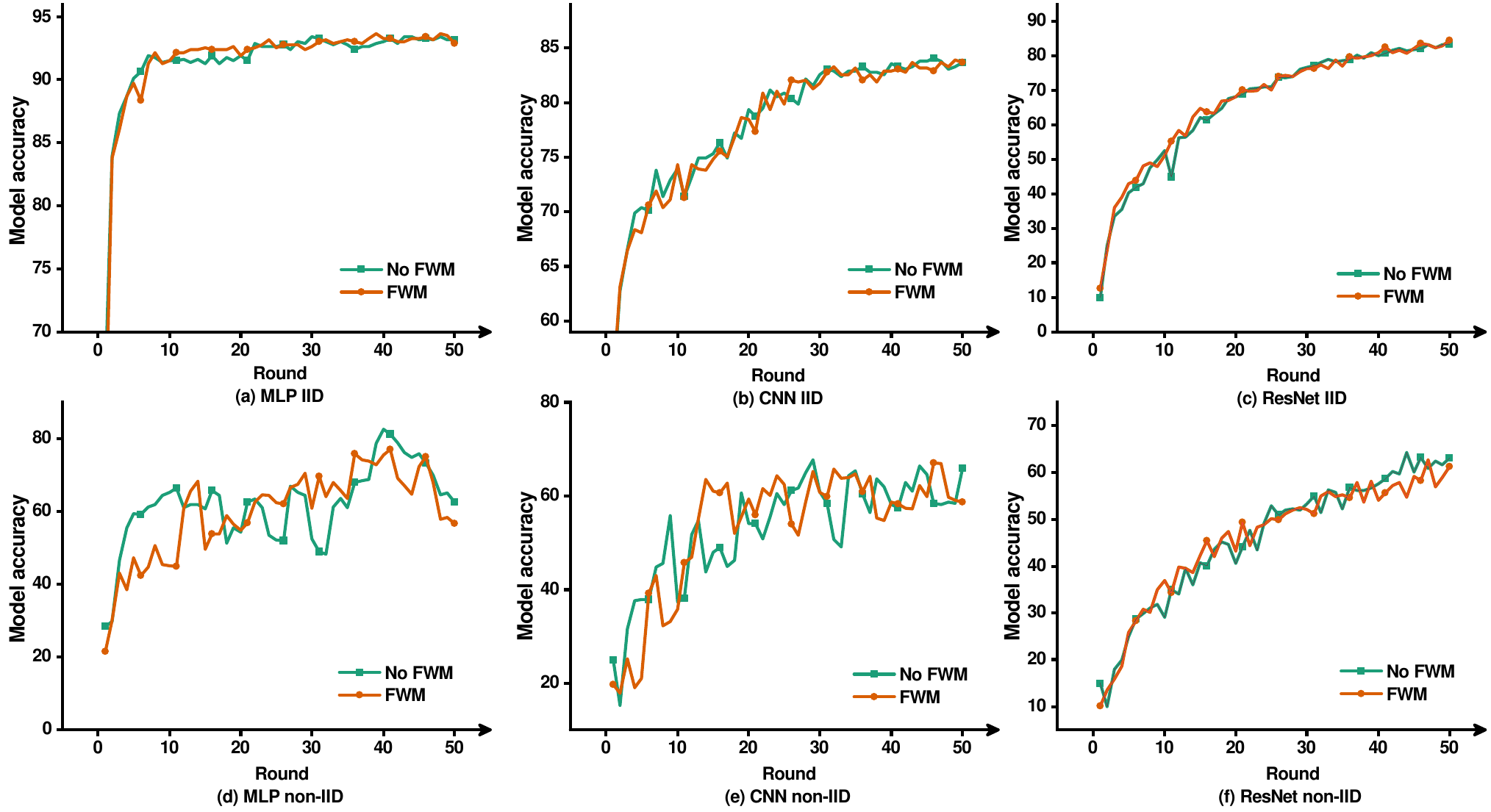}
  \caption{Impact of federated watermarking on model accuracy.}
  \Description{Impact of federated watermarking on model accuracy.}
  \label{fig2}
\end{figure}

To evaluate the fidelity of FedCIP, we conducted experiments on three different models and datasets, namely MLP, CNN, and ResNet, using both iid and non-iid data distributions. We compared the accuracy of the global model with and without watermarking. The results are shown in Figure \ref{fig2}, where the accuracy of each model with different watermark lengths on the same dataset is displayed separately.

As depicted in Figure \ref{fig2}, we conducted experiments to test the effect of having or not having watermarks on model accuracy for different network structures on IID and non-IID datasets. The accuracy rates of MLP, CNN, and ResNet18 with and without watermark on the IID dataset are represented by a, b, and c, while d, e, and f represent the accuracies of MLP, CNN, and ResNet18 with and without watermark on the non-IID dataset. The horizontal axis in the figure shows the number of training rounds, while the vertical axis represents the accuracy rate. The green dash represents the accuracy without watermarks, while the orange line indicates the accuracy with watermarks. A comprehensive analysis of the experiments depicted in Figure \ref{fig2} reveals that the presence of a watermark has little impact on the accuracy of the model. Therefore, FedCIP has a negligible effect on the performance of the federated learning model.

\subsubsection{Reliability}

To evaluate the reliability of the watermark detection in FedCIP, we conducted experiments varying three parameters related to watermark embedding $T_i$, $\alpha$, and $\beta$.  $T_i$ mainly affects the rapid replacement of the watermark in every round when $T_i$=1, which can impact the watermark embedding. The experiments aimed to examine the effect of different values of 
$T_i$, $\alpha$, and $\beta$ on the watermark detection rate. Specifically, we measured the watermark detection accuracy, which reflects the similarity between the watermark before the embedding and the watermark extracted from the suspicious model.

\begin{figure}[h]
  \centering
  \includegraphics[width=\linewidth]{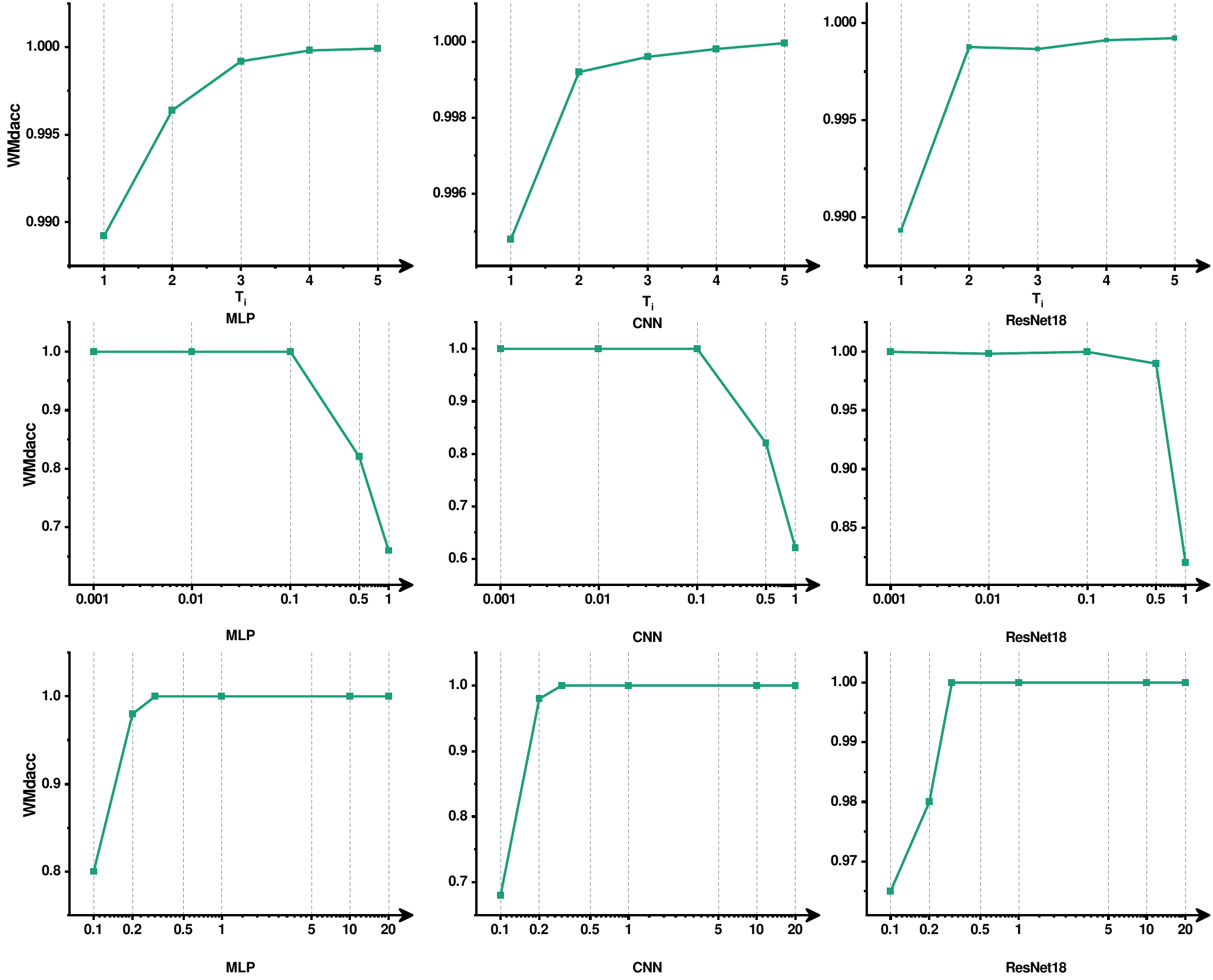}
  \caption{The effect of different watermark cycle $T_i$, watermark power $\alpha$, and $\beta$ on the watermark detection accuracy.}
  \Description{The effect of different $T_i$, $\alpha$, and $\beta$ on the watermark detection accuracy.}
  \label{fig3}
\end{figure}

The results of the reliability of the watermark are presented in Figure \ref{fig3}. The vertical coordinate represents the watermark detection accuracy in the average of 50 detections. The first row of the figure shows the effect of $T_i$ on the watermark detection accuracy. It can be seen that the watermark detection accuracy remains high $(99\%)$ even when $T_i$ is set to 1, meaning that the watermark is rapidly replaced in each round. As $T_i$ increases, the watermark detection accuracy approaches $100\%$.

The second row of Figure \ref{fig3} shows the impact of the watermark strength parameter $\alpha$ on the watermark detection accuracy. The results indicate that a value of $\alpha$ between 0.001 and 0.1 produces $100\%$ watermark detection accuracy. However, when $\alpha$ increases to around 0.5, the watermark detection accuracy rapidly decreases. This decrease occurs because large $\alpha$ values cause the weights to become too large, which in turn reduces the success rate during the next watermark replacement.

The third row of Figure \ref{fig3} shows the impact of the watermark strength parameter $\beta$ on the watermark detection accuracy. Unlike $\alpha$, smaller $\beta$ values (e.g., 0.1) result in a lower success rate for watermark detection. As $\beta$ increases (between 0.5 and 20), the watermark detection accuracy increases to $100\%$. The value of $\beta$ affects the first round in $T_i$, during which the watermark is replaced with a new one. A smaller $\beta$ value means that the new watermark is weaker, resulting in a lower success rate for watermark replacement.

\subsubsection{Capacity}

In this section, we will test the length of the embeddable watermark. We will keep increasing the length of the watermark and test its effect on the accuracy of the model for the same number of rounds.

\begin{figure}[h]
  \centering
  \includegraphics[width=\linewidth]{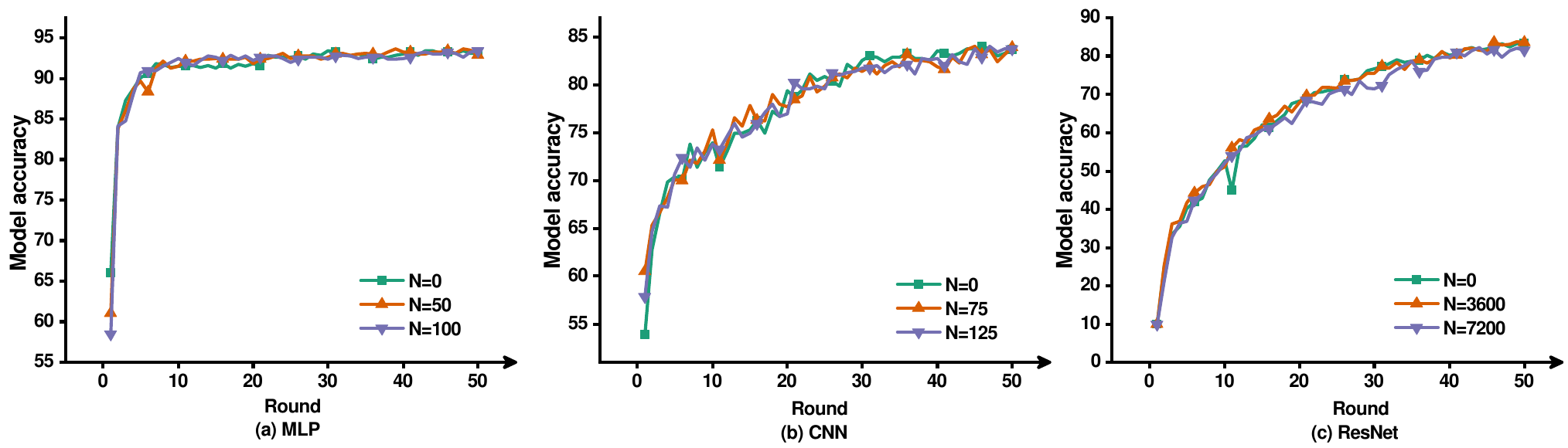}
  \caption{The effect of different lengths of watermarks on the model accuracy.}
  \Description{The effect of different lengths of watermarks on the model accuracy.}
  \label{fig4}
\end{figure}

Figure \ref{fig4} shows the effect of different watermark lengths on the accuracy of the model. N is the length of the watermark. From the figure, it can be seen that increasing the length of the watermark within a certain range does not significantly affect the accuracy of the model.

\subsubsection{Robustness}

A good watermarking scheme can make the watermark still detectable even when subjected to watermark removal attacks such as pruning \cite{32} and fine-tuning \cite{12,31}. Pruning refers to the removal of unimportant parameters from the model by the model stealer after stealing the model. This approach is able to corrupt the backdoor hidden in inconspicuous parameters while keeping the model usable. Specifically, after getting the model, the thief sorts the model weights in absolute magnitude and sets the weights near 0 to 0. Fine-tuning is where the model thief continues to train the model using the same dataset as the model training features. This method is able to modify the weights inside the model while ensuring that the model is available for the purpose of corrupting the potential watermark.

We use these two methods separately to attack our watermarking scheme with the following results.

\begin{figure}[h]
  \centering
  \includegraphics[width=\linewidth]{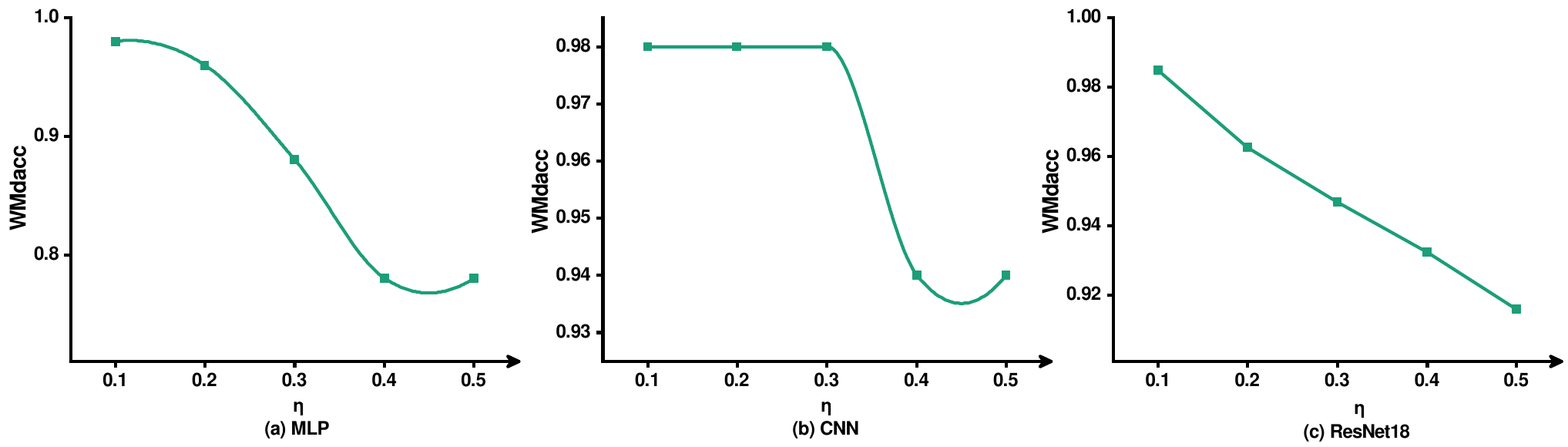}
  \caption{The effect of different purning rate $\eta$ on the watermark detection accuracy.}
  \Description{The effect of different purning rate $\eta$ on the watermark detection accuracy.}
  \label{fig5}
\end{figure}

As shown in Figure \ref{fig5}, we evaluated the resistance of FedCIP against pruning attacks. The horizontal axis represents the pruning coefficient, defined as the proportion $\eta$ of the weight value near 0 in the model that is set to 0. The vertical axis represents the watermark detection accuracy. The results show that the impact on watermark detection accuracy is within $5\%$ for a pruning coefficient no greater than 0.5. However, it is important to note that the pruning coefficient should not be set too large, as it can negatively affect the accuracy of the model. These results demonstrate that FedCIP is able to maintain its watermark detection accuracy even in the face of pruning attacks, further validating its robustness against attacks on model weights.

\begin{figure}[h]
  \centering
  \includegraphics[width=\linewidth]{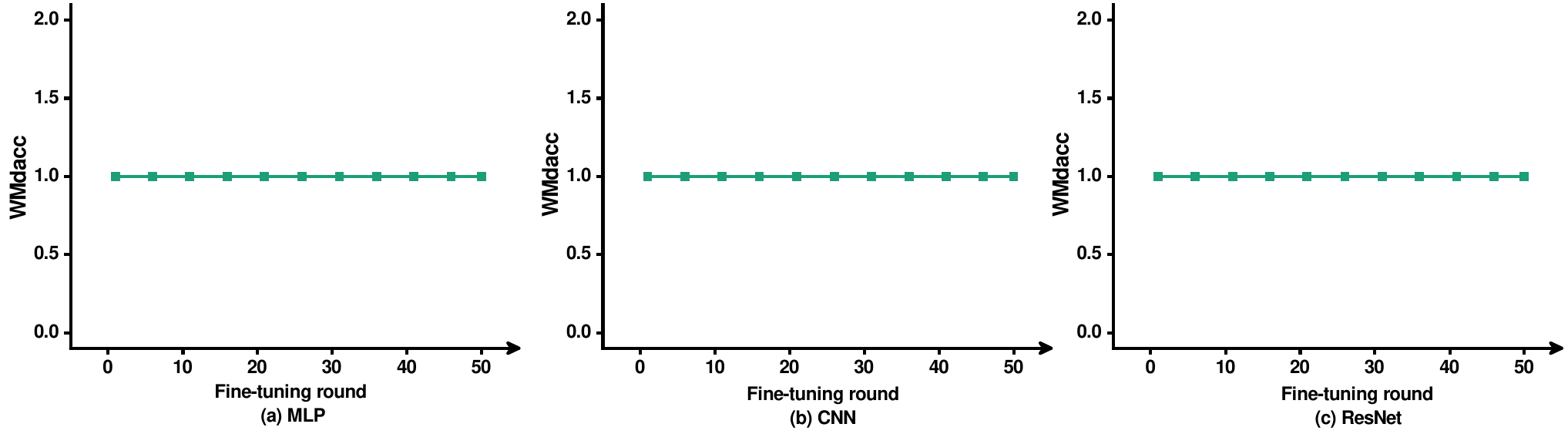}
  \caption{The effect of different fine-tuning rounds on the watermark detection accuracy.}
  \Description{The effect of different fine-tuning rounds on the watermark detection accuracy.}
  \label{fig6}
\end{figure}

The fine-tuning experiment aims to test the effectiveness of FedCIP against fine-tuning attacks. The results are shown in Figure \ref{fig6}, where the horizontal coordinate represents the number of fine-tuning rounds and the vertical coordinate represents the watermark detection accuracy. The experiment shows that even after 50 rounds of fine-tuning, the watermark detection accuracy remains at $100\%$. This indicates that FedCIP is resistant to fine-tuning attacks and can effectively protect the intellectual property of the federated learning model.

\section{Discussion}

The approach proposed in this paper is a novel approach to federated learning as it addresses the issue of tracking traitors in a privacy-preserving manner. The ability to track traitors in federated learning is crucial in ensuring the security of the learning process, and FedCIP achieves this goal while being compatible with parametric encryption schemes. Although the method of tracking the traitor in this paper depends on the number of times the traitor itself stole the model, it increases the cost of the traitor compared to previous approaches that do not have the ability to track the traitor. This means that the approach in this paper puts pressure on the traitor, making it more difficult for them to steal models with impunity.

However, the parameters used in the FedCIP approach, $\alpha$, and $\beta$, are conventional parameters and not optimized for the model. For future work, we can explore designing better parameter schemes that have less impact on the model and can enhance the capacity of the watermarking technique.

Overall, the FedCIP approach presented in this paper is a step forward in addressing the challenges faced by federated learning systems in terms of security and privacy. It provides a practical and effective way to track traitors, ensuring the security of the federated learning process.

\section{Conclusions}

In this paper, we explore the challenges of protecting intellectual property rights in federated learning and review existing research on watermarking techniques from both the client and server sides. We identify the limitations of each approach and propose a new client-side watermarking approach called FedCIP that allows for traitor tracking while maintaining compatibility with federated learning security aggregation such as parameter encryption. We provide a detailed description of the design and implementation of FedCIP and evaluate its performance in terms of model accuracy and watermark detection accuracy using three different models and datasets. Our results demonstrate that FedCIP is effective in protecting intellectual property in federated learning while also adding pressure on traitors to reduce the risk of IP theft. Future work could focus on optimizing the parameter schemes for FedCIP to minimize its impact on model performance.

\balance

%
\bibliographystyle{ACM-Reference-Format}
\bibliography{references}

\end{document}